\documentstyle[twocolumn,aps]{revtex}
\begin{document}
\tightenlines
\preprint{GRP-463}
\draft
\title{ 
The central density of neutron stars in close binaries}
\author{Alan G. Wiseman}
\address{Theoretical Astrophysics, 130-33,
California Institute of Technology\\
Pasadena, CA 91125\\
{\rm E-mail: agw@tapir.caltech.edu}}
\date{\today} 
\maketitle 
\begin{abstract}
Recent numerical simulations of coalescing binary neutron stars
conducted by Wilson, Mathews and Marronetti (WMM)
show a rising central energy density of the stars
as the orbital separation shrinks,
{\it i.e.} the stars are individually crushed as they near coalescence.
They claim this ``star-crushing'' effect is partially due to a 
non-linear, first post-Newtonian order enhancement
of the self-gravity of each star caused by the presence of the other star.
We present a concrete calculation which shows, within general relativity,
{\it first post-Newtonian order interactions with the other star
leave the central energy density unchanged as the orbital radius shrinks}.
The results presented here are in sharp disagreement with the WMM claim.
However, alternative gravitational theories,
such as Brans-Dicke theory,
can exhibit a small crushing effect in the binary constituents
as they near coalescence.
We show that the absence of the star-crushing effect at 
first post-Newtonian order
is related to adherence to the strong equivalence principle.
\end{abstract}


Binary neutron stars spiraling toward their final
coalescence under the dissipative influence of gravitational 
radiation reaction forces are the primary sources
gravitational waves which might be detected
by interferometric gravitational wave
detectors such as LIGO and VIRGO\cite{Abramovici92,Thorne95}.
Extracting the gravitational waves from the detector noise and 
using the information encoded in the signals requires a
precise knowledge of the expected waveforms \cite{Thorne95}.
The primary method for studying the behavior of these
highly relativistic
binaries has been the ``post-Newtonian'' approximation
carried to high relativistic order
\cite{tegp,bdiww,ww,pnexp}.
Recently, Wilson, Mathews and Marronetti (WMM) \cite{wmm}
have used numerical hydrodynamic simulations as an
alternative method for studying the
late stages of binary neutron-star inspiral.
These simulations, 
based on the simplifying assumption that space-time
is spatially conformally-flat (SCF) \cite{scf},
obtain a sequence of equilibrium configurations
of the neutron stars at closer and closer orbital separation.
Their numerical results 
show that the neutron stars are individually compressed
and the central densities begin to rise
as the orbital separation is decreased.
[See Mathews and Wilson \cite{wmm} Figure 2.]
Such a crushing-effect has never been reported in post-Newtonian
calculations.
This discrepancy forces a re-examination of the validity of both the 
post-Newtonian calculations and the numerical simulations.
(See Eardley \cite{eardley} for a discussion.)

It is tempting to simply dismiss this unusual
star-crushing effect as an artifact of the 
SCF constraint used in the WMM numerical simulations.
However, WMM further claim that the star-crushing effect is partially due
to a non-linear first post-Newtonian order (1PN) enhancement of the gravitational
potential.
(See Mathews and Wilson \cite{wmm} Appendix A:
``... the [first] post Newtonian expansion does indeed exhibit
an enhancement to the gravitational potential ...".)
However, the 1PN metric is naturally spatially conformally flat;
therefore such an effect would be independent of the SCF assumption.
[See Eq.(\ref{metric}).]
It is also tempting to dismiss the WMM claim on the
grounds that Newtonian tidal-effects will stretch the individual
stars, and thereby reduce the central densities of stars \cite{lai}.
However, WMM have made it clear that their proposed crushing mechanism
will swamp -- by a factor of $10^4$ --
the rather small reduction in central density due to
tidal distortion. [See Mathews and Wilson \cite{wmm} Eq.~(20).]

There is actually a substantial body of literature
on post-Newtonian effects for extended bodies 
in binaries that seemingly would have discovered a
1PN star-crushing effect if it were real. 
(See {\it e.g.} \cite{magnum,bs}.)
There are also published arguments that show that
enhancements of the local gravity due the
presence of other gravitating bodies involve a 1PN
violation of the strong equivalence principle.
(See Will \cite{tegp} Chapter 6.3.)
However, WMM specifically  claim a 1PN origin of the star-crushing effect;  
therefore we offer a specific rebuttal to the claim.

In this paper we present a brutally straightforward,
first post-Newtonian order calculation of the central energy density of a
compact star in a close binary.
We solve the 1PN hydrodynamic equation
(including non-linear gravitational interactions with the other star)
for the density profile of the star.
At intermediate steps in the calculation,
relativistic corrections appear which give the illusion that each
star is being compressed by the presence of the other star.
In the end, for a first post-Newtonian expansion of general relativity,
all the relativistic corrections involving the other star cancel,
and the true physical quantities
(the central energy density, and the proper radius)
are unaffected by the presence of the other star.
Thus WMM's claim of a 1PN origin of
the star-crushing effect is false.
The confusion can be summed up as follows:
the potential [in this case $(g_{00}+1)]$
is made ``deeper'' by non-linear relativistic factors involving the other star;
thus, when solving for the density profile, 
it appears that the star is being compressed by the presence of the other star.
However, the relativistic factors also re-enter the  calculation
as 1PN corrections to the spatial part of the metric.
The spatial metric components alter the spatial line element 
in the calculation of the
proper radius of the star, and they alter the 
proper volume element in the calculation of the conserved baryon number. 
In general relativity, at first post-Newtonian order, these effects
exactly cancel, and the physical quantities are left unaffected by the 
other star.  


We consider two perfect fluid blobs (stars) with
baryon masses $M_{bA}$ and $M_{bB}$.
For simplicity we assume that the stars are in a nearly circular orbit
with coordinate radius $R_o$.
We boost to a comoving, post-Newtonian coordinate system 
where the fluid in star-A is {\it momentarily} at rest,
and the origin instantaneously coincides with the center of star-A.
[Star-A is still being accelerated toward star B.]
In this moving frame the parameterized, post-Newtonian  (PPN)
metric coefficients are
\cite{tegp,pnexp2}
\begin{mathletters}
\label{metric}
\begin{eqnarray}
&& g_{00} = -1 + 2 U({\bf x},t) - 2 \beta U({\bf x},t)^2
+ 2(\gamma+1) \Phi_1 ({\bf x},t)  \nonumber \\
&& + 2(3\gamma-2\beta +1) \Phi_2 ({\bf x},t)
+ 2 \Phi_3 ({\bf x},t)  
+ 6\gamma \Phi_4 ({\bf x},t)  \\
&& g_{ij} = [1 + 2 \gamma U({\bf x},t)] \delta_{ij}  \\
&& g_{0j} = -(1/2)(4 \gamma + 3) V_j ({\bf x},t) -(1/2)W_j ({\bf x},t) \; ,
\end{eqnarray}
\end{mathletters}
where $\beta$ and $\gamma$ are the PPN
coefficients which characterize alternative theories of gravity.
Solar system tests have constrained their values to be very 
close to unity.
In general relativity $\beta \equiv \gamma \equiv 1$.
Also notice the metric is spatially conformally flat (SCF) \cite{scf}.
The sources for the potentials (the baryon density $\rho$, pressure $P$,
and internal thermal energy $\rho \Pi$)
are measured in the comoving frame of the fluid.

We now define and evaluate the potentials appearing in Eq.(\ref{metric}).
(Details can be found in \cite{tegp}.)
The Newtonian potential is
\begin{eqnarray}
U({\bf x},t&& ) \; \equiv \int { \rho( {\bf x^\prime},t )  d^3 {\bf x^\prime }
\over | {\bf x} - {\bf x^\prime} | } 
= { M_{bB} \over R_o} + {M_{bB} {\bf R_o \cdot \ x} \over R_o^3 }
\nonumber \\
&&+ 4\pi \biggl [ {1 \over r} \int_0^r \rho_A(r') \; {r^\prime}^2  d r^\prime 
 + \int_r^{L}  \rho_A(r')  {r^\prime}  d r^\prime \biggr ] \,. 
\label{potU}
\end{eqnarray}
where $L$ denotes the -- as yet unknown -- coordinate radius of the star,
${\bf R_o}$ is the vector separation of the body centers,
$R_o=|{\bf R_o}|$ is the constant coordinate orbital separation,
${\bf x}$ is the field point, $r=|{\bf x}|$ is the
radial coordinate measured from the center of star-A.
The second expression in Eq.(\ref{potU})
is valid only within the quasi-stationary star-A.
We have also neglected tidal terms. 
The $({\bf R_o \cdot x})$-term gives rise to the bulk acceleration
of star-A toward star-B.  
Over the extent of star-A, it has zero average radial ($r$) component;
therefore it cannot contribute to the compression. 
We will neglect this term and other terms which clearly cannot contribute
to the radial compression.
The Newtonian potential can now be written
\begin{equation}
U(r<L) = { M_{bB} \over R_o } + U_A (r) \; ,
\label{potUcompact}
\end{equation}
where $U_A(r)$ is defined as the ``$4\pi$''-term in Eq. (\ref{potU}).
For future reference, we compute the radial derivative
\begin{equation}
U,_r =  - { 4\pi \over r^2 } \int_0^r \rho(r') \; {r^\prime}^2 \; d r^\prime
\equiv - { M_{bA}(r) \over r^2 } \; .
\label{potUr}
\end{equation}
The potential $\Phi_1$ is the post-Newtonian contribution 
from the kinetic energy.
As we have boosted to a frame where star-A is a rest, there is no
bulk motion of the fluid in star-A to contribute to this term.  
We also neglect internal fluid motion in star-A,
as this actually decreases the central density \cite{lai}. 
Thus we have
\begin{eqnarray}
&& \Phi_1({\bf x},t) = \int_B { v^2 \rho( {\bf x^\prime},t )  
d^3 {\bf x^\prime }
\over | {\bf x} - {\bf x^\prime} | } = v_B^2 {M_{bB} \over R_o}  \; ,
\label{phi1r}
\end{eqnarray}
where $v_B$ is the velocity of star-B measured in the frame where
star-A is momentarily at rest.
The potential $\Phi_2$ is the post-Newtonian contribution 
from the gravitational energy
\begin{equation}
\Phi_2({\bf x},t) \equiv 
\int { U({\bf x^\prime},t)\rho( {\bf x^\prime},t )  d^3 {\bf x^\prime }
\over | {\bf x} - {\bf x^\prime} | } \; .
\label{phi2}
\end{equation}
Proceeding as with previous terms we have
\begin{equation}
\Phi_2,_r =  - {M_{bB} \over R_o} { M_A(r) \over r^2 }
- { 4\pi \over r^2 } \int_0^r \rho_A(r') U_A(r') \; {r^\prime}^2 \; 
d r^\prime \; .
\label{phi2r}
\end{equation}
The origin of the WMM star-crushing claim is that the first term in 
Eq.(\ref{phi2r}) will conspire with the Newtonian gravitational pull
in Eq.(\ref{potUr}) to enhance the compression of
star-A.
Noting that the WMM-factor $(W^2-1) \sim M_{bB}/R_o$, 
we see that radial enhancement to the Newtonian compressional 
force appearing in Eq.(\ref{phi2r}) is
the force related to the potential energy
shown in Mathews and Wilson \cite{wmm} Eq.(19)
[$\Delta E_{GR} \approx 2 (W^2-1) G M^2/L$].
Thus, we are unambiguously addressing the term that 
WMM claim is the origin of the crushing-effect.

The potentials  $\Phi_3$  and $\Phi_4$ are similar to $\Phi_1$ with the 
kinetic energy replaced by $\rho \Pi$ and $P$ respectively.
Finally, the potential can be evaluated using formulae in
\cite{tegp} The results: $V_j=M_{bB} v_B^j/R_o$, and $W_j=0$ for circular 
orbits.

We now look at at the spatial components of the Euler equation
in our instantaneously comoving-moving frame of star-A.
In this frame, the fluid is 
momentarily stationary ($u^k=P_{,t}=0$ for star-A), 
and we can write
\begin{equation}
{ dP \over d r } = - (\rho + \rho \Pi + P) 
\biggl [ { d \over dr }  \log \sqrt {-g_{00}}  \biggr ] \; .
\end{equation}
We have neglected a number or second post-Newtonian terms, 
as well as terms which do not contribute to the radial compression.

Expanding the logarithm, substituting the potentials,  
and discarding second post-Newtonian terms, we have
\begin{eqnarray}
{ dP \over dr } - \rho U,_r = && \epsilon \rho \biggl [
2(1-\beta)U U,_r + (3 \gamma - 2 \beta +1) \Phi_2,_r \nonumber \\
&& \;\; + \Phi_3,_r
+ 3 \gamma \Phi_4,_r
+ [\Pi + (P/\rho)] U,_r \biggr ] \; .
\label{hydrostat}
\end{eqnarray}
On the left we have the familiar pressure gradient
balancing the Newtonian gravitational attraction.
On the right we have all the first post-Newtonian corrections.
We have inserted a dummy factor of $\epsilon (=1)$ to keep track of
post-Newtonian terms.
Multiplying by $r^2/\rho$, differentiating, and using the expressions
for the radial derivatives of the potentials,
we have 
\begin{eqnarray}
&& {1 - \epsilon(3\gamma-4\beta+3)(M_{bB}/R_o) \over 4 \pi }
{1 \over r^2} {d \over dr} 
\biggl [ {r^2 \over \rho} {dP_A \over dr} \biggr ] + \rho_A  \nonumber \\
&& = -\epsilon \biggl \{ {\beta - 1 \over 2 \pi r^4 }  M_A(r)^2
+ \rho_A \left [ (3 \gamma-4\beta +3) U_A + \Pi_A \right ]
\nonumber \\
&& \;\;\;\;\;\;\;\;\; + (3\gamma+1) P_A 
+ {d \over dr} 
\biggl ( \Pi_A+ { P_A \over \rho_A } \biggr ) { M_A(r) \over 4\pi r^2 }
\biggr \} \; .
\label{sloppy}
\end{eqnarray}
We have discarded terms of second post-Newtonian order, 
{\it i.e.} $O(\epsilon^2)$.
On the left of Eq.(\ref{sloppy})
we have Newtonian terms, as well as the post-Newtonian terms related to the
presence of star-B.
On the right we have only post-Newtonian self-interaction terms.
The terms on the right will affect the matter configuration of star-A,
but they would do so whether or not star-B is nearby.
Our focus is on deriving the 1PN effect star-B has
on the matter configuration in star-A.
This can be done by setting the right hand side of Eq.~(\ref{sloppy})
to zero and solving the homogeneous equation.
In a full solution to Eq.(\ref{sloppy})
any non-linear interplay between the post-Newtonian
corrections on the left and right sides of Eq.(\ref{sloppy}) 
would be $O[\epsilon^2]$, and could be neglected. 
Throughout the remainder of this calculation, we
ruthlessly discard post-Newtonian self interaction terms.
(A 1PN solution of Eq.(\ref{sloppy})
-- including the self interactions on the right side -- 
will be presented elsewhere \cite{prep1}.)

Substituting the polytropic equation of state
\begin{equation}
P = K \rho^{1+1/n} \; ,
\label{polytrope}
\end{equation}
into the homogeneous equation, and defining the quantities
\begin{mathletters}
\label{quantities}
\begin{eqnarray}
\rho &&= \rho_c \theta^n \\
a &&= \sqrt{ {(n+1)K \over 4 \pi} }  \\
\lambda&&= a \rho_c^{{1-n \over 2n}}  
\biggl [ 1-{\epsilon \over 2} {M_{bB} \over R_o} (3\gamma-4\beta +3)\biggr ] \\
\xi &&= r/\lambda \;.
\end{eqnarray}
\end{mathletters}
we obtain the dimensionless Lane-Emden \cite{chandra} equation
\begin{equation}
{1 \over \xi^2} {d \over d\xi} 
\biggl [ \xi^2  {d\theta \over d\xi} \biggl ] + \theta^n = 0 \;.
\label{laneemden}
\end{equation}
[We restrict our attention to non-singular ($n<3$) solutions \cite{chandra}.]
The solutions  $\theta(\xi;n)$ of this non-linear differential equation
must satisfy the boundary conditions
\begin{eqnarray}
\theta(0;n) = 1 \; \; , \; \;\;\;\; \theta^\prime(0;n) = 0 \;.
\label{bc}
\end{eqnarray}
The first condition defines $\rho_c$ as the central baryon density.
The second insures there is no pressure gradient at the origin.
The density profile of star-A can then be written
\begin{equation}
\rho_A (r) = \rho_c \, \theta(r/\lambda ;n)^n \;.
\label{densprofile}
\end{equation}
Notice that $\rho_A(r)$ has an apparent 1PN dependence on 
the potential produced by the other star ($M_{bB}/R_o$)
through Eq.(\ref{quantities}c).
The coordinate radius $L$ of the stellar surface is the value of 
$r$ where the pressure (and thereby the density) vanishes;
it is given by
\begin{equation}
L = \lambda \, \xi_{1n} \; ,
\label{coordrad}
\end{equation}
where $\xi_{1n}$ is the smallest positive root satisfying 
$\theta(\xi_{1n};n)=0$.

The total baryon mass can be obtained by integrating
\begin{mathletters}
\label{mbfinal}
\begin{eqnarray}
M_{bA} && = \int_A \rho \; d({\rm proper \;\; volume})  
 = \int_A \rho u^0 \sqrt{ - g} d^3x \\
&& = \int_A \rho [1+3 \epsilon \gamma U ] d^3x \\
&& = 4 \pi a^3 f(n) \rho_c^{3-n \over2n}
\biggl [ 1+{3 \over 2} {M_{bB} \over R_o} (4\beta-\gamma-3)\biggr ]  \\
&& = M_{bA}\left [ K,n,\rho_c,(4\beta-\gamma-3)(M_{bB}/ R_o) \right ] \\
&& = 4 \pi a^3 f(n) \rho_c^{3-n \over2n}
 \;\;\; ( {\rm in \;\; gen. \;\; rel.\;\;}\beta=\gamma=1  ) \;, \\
&& = M_{bA}[ K,n,\rho_c]
 \;\;\;\; ( {\rm in \; general \;\; relativity \;\;} ) \;,
\end{eqnarray}
\end{mathletters}
where $f(n) = \xi_{1n}^2 | \theta^\prime ( \xi_{1n} , n ) | $ is 
solely a function
of the polytropic index $n$.
We have discarded self-interaction terms
and 2PN terms.
(We have also set the dummy $\epsilon=1$.)
{\it Notice, -- and this is the point of the paper -- 
in general relativity at 1PN order
for a star with fixed baryon mass ($M_{bA}$)
and a fixed equation of state ($K$ and $n$),
there will be no change in the
central density as the binary orbital separation changes.}
The functional dependence on $M_{bB}/R_o$ cancels out of Eq.~(\ref{mbfinal})
for general relativity (or any gravitational theory with $4\beta-\gamma-3=0$).
The inclusion of the self interactions that we have 
been discarding would complicate the functional
form shown Eq.~(\ref{mbfinal}e), but it would not affect the
the functional dependence depicted in Eq.~(\ref{mbfinal}f);
therefore our 
{\it no-crushing} conclusion
would  hold even if we were to include 
the self interactions on the right of Eq~(\ref{sloppy}).

The proper radius, $\Delta_A$, of the star can be obtained by integrating
\begin{mathletters}
\label{properrad}
\begin{eqnarray}
\Delta_A&& = \int_0^{L} \sqrt{ g_{rr} } dr \\
&& = L_{N}
\biggl [ 1+{n \over 3-n} {M_{bB} \over R_o} (4\beta-\gamma-3)\biggr ] \\
&& = L_{N}  \;\;\;\; ( {\rm in \;\; general \;\; relativity \;\;} 
\beta \equiv \gamma \equiv 1 ) \;.
\end{eqnarray}
\end{mathletters}
where 
\begin{equation}
L_{N} = a \, \xi_{1n} \,
\biggl ( {M_{bA} \over 4 \pi a^3 f(n) } \biggr )^{1-n\over3-n}
\end{equation}
is the radius of a the isolated Newtonian star with baryon mass $M_{bA}$.
{\it Notice, in general relativity at 1PN order, 
the proper radius 
of the star in a close binary is unaffected by the presence of the other star.}

In the Brans-Dicke theory of gravity $\beta \equiv 1$ and
$|\gamma-1|\lesssim 0.001$ \cite{tegp}; therefore the compression
effect illustrated in Eq.  (\ref{mbfinal}) and (\ref{properrad})
will be small. However, even with these constrained values of
$\beta$ and $\gamma$, this effect is potentially larger than those
due to tidal distortion. [See Mathews and Wilson \cite{wmm} Eq.(20).]

The 1PN crushing effect we have been discussing  can be interpreted as
a change in the local value of Newton's gravitational constant
due to the presence of the other star in the binary.
That is, Eq.~(\ref{mbfinal}c) and Eq.~(\ref{properrad}b) can be obtained
from the standard Newtonian formula \cite{chandra}  by the replacing
$G=1$ with
\begin{equation}
G_{local} =  1 - (4\beta - \gamma - 3)(M_{bB} /R_o)  \; .
\end{equation}
Such effects arise in gravitational theories that violate 
the strong equivalence principle.  [See reference \cite{tegp} Eq.~(6.75).]


\medskip
Notice, we have only shown that there is no crushing effect at 1PN order in
general relativity.
As the conventional (pefect fluid) post-Newtonian expansion of the metric
is naturally spatially conformally flat at 1PN order,
we cannot conclude with certainty
that the crushing effect that WMM see is 
an artifact of the SCF constraint used in the numerical simulations.
Indeed, it is possible that the crushing effect is real, 
and WMM's only sin was to incorrectly assign a 1PN origin to the effect.
The question remains whether the SCF boundary condition
is somehow responsible for (or amplifies) the crushing effect
in the simulations \cite{altern}.

The post-Newtonian expansion of the metric is not 
spatially conformally flat at second post-Newtonian order \cite{cliff,reith}; 
therefore a thorough 2PN investigation 
may clarify the difference between the conventional post-Newtonian 
analysis of the inspiral problem and the SCF numerical analysis \cite{prep1}.
WMM claim that the SCF contraint is justified, 
because it removes only the radiation from the solution.
This is not true. 
Radiation only enters the metric at 2.5PN order;
thus altering the metric at 2PN order by imposing the SCF constraint
will change the longitudinal fields that govern the periodic orbital 
motion and the individual body configurations.
Although it is true that the WMM equilibrium solutions are
exact solutions (up to numerical accuracy) of the Einstein
field equations,
they are solutions which differ from the actual inspiral
metric at the 2PN level.
In this formal sense, the SCF simulations are only as accurate 
as a 1PN simulation.


It is a pleasure to thank Patrick Brady, Scott Hughes, Lee Lindblom,
Katrin Shenk and Clifford Will for their helpful insights.
This work was supported by NSF grant PHY-9424337 and AST-9417371.




\begin{references}
\bibitem{scf} 
For the uninitiated, spatial conformal flatness
means the spatial part of the metric can be written
as $g_{ij} = (conformal \;\; factor) \times \delta_{ij}$.
\bibitem{pnexp2} 
Virtually all Newtonian gravitational effects can be obtained
from the metric coefficients $g_{00}=-1+2U$, $g_{0i}=0$,
and $g_{ij}=\delta_{ij}$.
We are including ``post''-Newtonian (1PN) corrections to all the
metric coefficients.
\bibitem{altern} 
Keep in mind that the imposition
of the SCF constraint
is not an ``alternative theory of gravity''
(in the sense that Brans-Dicke is an alternative theory).
Rather, the SCF constraint is really a stringent boundary condition
on the solutions to the actual general relativistic field equations.
\end{references}
\end{document}